\documentclass[aip,jcp,reprint,groupedaddress]{revtex4-2}%
\usepackage{graphicx}
\usepackage{tabularx}
\usepackage{dcolumn}
\usepackage{longtable}
\usepackage{tensor}
\usepackage{placeins}
\usepackage{bm}
\usepackage{xcolor}
\usepackage{amsmath}
\usepackage{amsfonts}
\usepackage{amssymb}
\usepackage{float}%
\usepackage[english]{babel}
\providecommand{\U}[1]{\protect\rule{.1in}{.1in}}

\begin{document}

\preprint{AIP/123-QED}

\title{Ehrenfest dynamics accelerated with SPEED}
\author{Alan Scheidegger}
\email{alan.scheidegger@epfl.ch}
\author{Ji\v{r}\'{\i} Van\'{\i}\v{c}ek}
\email{jiri.vanicek@epfl.ch}
\affiliation{Laboratory of Theoretical Physical Chemistry, Institut des Sciences et
Ing\'enierie Chimiques, Ecole Polytechnique F\'ed\'erale de Lausanne (EPFL),
CH-1015, Lausanne, Switzerland}
\date{\today}

\begin{abstract}
Mixed quantum-classical methods, such as surface hopping and Ehrenfest dynamics, have proven useful for describing molecular processes involving multiple electronic states. These methods require propagating many independent trajectories, which is computationally demanding.
Therefore, we propose the single potential evaluation Ehrenfest dynamics (SPEED), a variation of Ehrenfest dynamics where all trajectories are propagated using a common local quadratic effective potential in the diabatic representation. This approach replaces the computational cost of propagating multiple trajectories with the evaluation of a single Hessian at each time step. We demonstrate the equivalence of standard Ehrenfest dynamics and SPEED in two realistic systems with (at most) quadratic diabatic potential energy surfaces and couplings: a quadratic vibronic coupling Hamiltonian model describing internal conversion in pyrazine and a model of atomic adsorption on a solid surface.
The efficiency gain of our approach is particularly advantageous in on-the-fly \textit{ab initio} applications. For this reason, we combined SPEED with the ALMO(MSDFT2) electronic structure method, which provides the diabatic potential describing charge transfer between two molecules. We find that SPEED qualitatively captures the temperature dependence of the hole transfer rate between two furan moieties and accurately predicts the final charge distribution after the collision. In contrast, but as expected, our approach is insufficient for describing photoisomerization of retinal due to the high anharmonicity of the potential energy surfaces.
\end{abstract}
\maketitle

\graphicspath{{./figures/}{C:/Users/Jiri/Dropbox/Papers/Chemistry_papers/2024/SPEED/figures/}{"C:/Users/GROUP LCPT/Documents/Alan/SPEED/figures/"}}

\section{Introduction}
The Born--Oppenheimer approximation separates the motion of electrons from the motion of nuclei. This separation leads to the introduction of the concept of a potential energy surface (PES) on which nuclei move~\cite{Born_Oppenheimer:1927}. However, many chemical processes involve multiple surfaces, requiring nonadiabatic molecular dynamics to capture the coupled dynamics of electrons and nuclei~\cite{Agostini_Curchod:2019}. Due to the exponential scaling of the exact grid-based solution, various approximate methods have been developed. Some approaches, such as the multi-configurational time-dependent Hartree method (MCTDH)~\cite{Beck_Meyer:2000}, still converge to the exact dynamics as the number of time-dependent basis functions approaches infinity.
However, a significant challenge for the MCTDH method lies in the need to construct a multistate model that accurately captures the PESs in important regions, such as in the vicinity of the Franck--Condon point or conical intersections~\cite{Penfold_Eng:2023}. A commonly used approach employs the vibronic coupling Hamiltonian model, which represents the PESs and their couplings as polynomial functions of nuclear coordinates.
Alternatively, on-the-fly evaluation of the potential can be used in combination with trajectory-based methods, where potential energy information is evaluated locally along trajectories and only when needed. The basis-function and trajectory-based approaches typically differ in their representations of electronic states. Since most electronic structure methods provide adiabatic states, the adiabatic representation is typically preferred for on-the-fly applications. In contrast, the diabatic representation is more convenient for most numerically exact quantum-mechanical methods. However, it is important to note that strict diabatic states, where all derivative couplings vanish, generally do not exist~\cite{Mead_Truhlar:1982}. As a consequence, the term ``diabatic states'' is typically used to refer to electronic states with negligible (altough nonzero) derivative couplings.

Ehrenfest dynamics~\cite{Ehrenfest:1927,Billing:1983,Li_Frisch:2005,Alonso_Jover-Galtier:2018,Kirrander_Vacher:2020} and surface hopping~\cite{Tully_Preston:1971,Tully:1990,Wang_Prezhdo:2016,Park_Shiozaki:2017,Zobel_Gonzalez:2021} are not only the most popular, but also the most efficient nonadiabatic dynamics methods.
Unlike surface hopping, Ehrenfest dynamics is occasionally employed as a single-trajectory method~\cite{Billing:1975,Ryabinkin_Izmaylov:2017a}. In this paper, ``Ehrenfest dynamics'' refers to the multi-trajectory version, unless otherwise specified. The single-trajectory approach is particularly useful for large systems with many electronic states, but its complete neglect of nuclear quantum effects limits its broader applicability.
In particular, a single Ehrenfest trajectory is insufficient for capturing internal conversion between electronic states belonging to different irreducible representations of the molecular point group~\cite{Scheidegger_Vanicek:2024}.
The significant limitations of single-trajectory Ehrenfest dynamics, along with the high computational cost of its multi-trajectory counterpart, highlight the lack of a nonadiabatic dynamics method that can capture early-time dynamics with the efficiency of a single Ehrenfest trajectory while remaining broadly applicable.
To address this, we propose an approach in which Ehrenfest trajectories are propagated using a shared local harmonic effective potential in the diabatic representation. While this concept can be applied to any method using trajectories to represent the nuclear wavefunction, here we focus on Ehrenfest dynamics, which is equivalent in the diabatic and adiabatic representations of the electronic states. This contrasts with surface hopping, which has shown better performance in the adiabatic representation~\cite{Tully:1998}.

\section{Theory}
In Ehrenfest dynamics of a system comprising $S$ electronic states, the $j$th trajectory is defined by its position $q_{j,t}$, momentum $p_{j,t}$ and electronic wavefunction $\mathbf{c}_{j,t}$ (an $S$-component vector of complex coefficients). In the diabatic representation, the three parameters are propagated according to the differential equations
\begin{align}
    \dot{q}_{j,t}&=m^{-1}\cdot p_{j,t},\\
    \dot{p}_{j,t}&=-\mathbf{c}_{j,t}^{\dagger}\mathbf{V}^{\prime}(q_{j,t})\mathbf{c}_{j,t},\\
    i\hbar\mathbf{\dot{c}}_{j,t} &=[T(p_{j,t})+\mathbf{V}(q_{j,t})]\mathbf{c}_{j,t}.
\end{align}
That is, at each time step, the $S\times S$ diabatic potential energy matrix $\mathbf{V}(q_{j,t})$ and its gradient $\mathbf{V}^{\prime}(q_{j,t})$ are necessary for each trajectory.
Instead, we propose using a common, time-dependent quadratic effective potential
\begin{equation}
\label{eq:Veff}
\begin{split}
    \mathbf{V}_{\mathrm{eff}}(q,t)&:=\mathbf{V}(q_{\mathrm{avg},t}) + \mathbf{V}^{'}(q_{\mathrm{avg},t})^{T} \cdot (q-q_{\mathrm{avg},t})\\
    &+(q-q_{\mathrm{avg},t})^{T} \cdot \mathbf{V}^{''}(q_{\mathrm{avg},t}) \cdot (q-q_{\mathrm{avg},t}) / 2
\end{split}
\end{equation}
corresponding to the second-order Taylor expansion of the potential $\mathbf{V}(q)$ around the average position
\begin{equation}
    q_{\mathrm{avg},t} = \frac{1}{N_{\mathrm{traj}}} \sum_{j=1}^{N_{\mathrm{traj}}}q_{j,t}.
\end{equation}
This new method, which we refer to as the ``single potential evaluation Ehrenfest dynamics'' (SPEED), is therefore defined by the set of differential equations
\begin{align}
    \dot{q}_{j,t}&=m^{-1}\cdot p_{j,t},\\
    \dot{p}_{j,t}&=-\mathbf{c}_{j,t}^{\dagger}\mathbf{V}_{\mathrm{eff}}^{\prime}(q_{j,t},t)\mathbf{c}_{j,t},\\
    i\hbar\mathbf{\dot{c}}_{j,t}&=[T(p_{j,t})+\mathbf{V}_{\mathrm{eff}}(q_{j,t},t)]\mathbf{c}_{j,t}.
\end{align}
From a computational perspective, the efficiency gain is evident, as only a single evaluation of the original, possibly anharmonic, or \textit{ab initio}, potential energy matrix $\mathbf{V}(q_{\mathrm{avg},t})$, its gradient $\mathbf{V}^{'}(q_{\mathrm{avg},t})$, and Hessian $\mathbf{V}^{''}(q_{\mathrm{avg},t})$ is required at each time step, regardless of the number of propagated trajectories.
In Sec.~\ref{sec:SI_computational_cost} of the Supplemental Material, we show that SPEED remains more efficient than Ehrenfest dynamics for systems where the number $D$ of nuclear dimensions does not exceed twice the number of trajectories necessary to achieve converged results.
We expect that the condition $D\le 2N_{\mathrm{traj}}$ is always met in practical applications.
In mixed quantum-classical simulations, the time step must be small enough to accurately capture the electronic motion. However, since the nuclei move more slowly than the electrons, the Hessian in Eq.~(\ref{eq:Veff}) can be updated less frequently to further improve efficiency. If the diabatic PESs are only weakly anharmonic, a single reference Hessian can be considered, as in the single-Hessian thawed Gaussian approximation~\cite{Begusic_Vanicek:2019}.
This would amount to replacing $\mathbf{V}^{''}(q_{\mathrm{avg},t})$ in Eq.~(\ref{eq:Veff}) with a constant, time-independent Hessian $\mathbf{V}^{''}(q_{\mathrm{ref}})$, but the potential $\mathbf{V}(q_{\mathrm{avg},t})$ and gradient $\mathbf{V}^{'}(q_{\mathrm{avg},t})$ are still time dependent.

SPEED has two special limits connecting it to multi- and single-trajectory Ehrenfest dynamics. First, when the diabatic PESs and couplings are at most quadratic, the effective potential~(\ref{eq:Veff}) becomes exact and SPEED is equivalent to multi-trajectory Ehrenfest dynamics. This property is further illustrated in Secs.~\ref{sec:ic_in_pyrazine} and~\ref{sec:adsorption}, where we demonstrate that SPEED and multi-trajectory Ehrenfest dynamics can be equivalent even in realistic systems. Second, when only a single trajectory is propagated in an arbitrary, anharmonic potential, SPEED reduces to single-trajectory Ehrenfest dynamics.

\section{Internal conversion in pyrazine}
\label{sec:ic_in_pyrazine}
\begin{figure}
\centering
\includegraphics[width=\columnwidth]{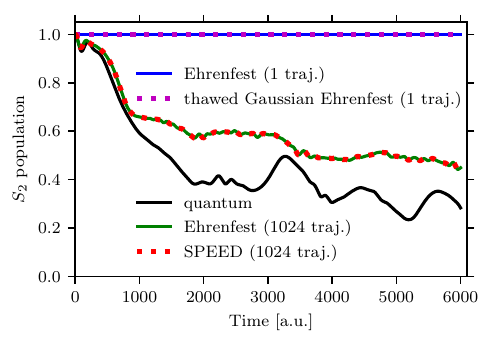}
\caption{Diabatic population of the second excited state $S_{2}(\pi\pi^{*})$ of pyrazine evaluated quantum mechanically and with various mixed quantum-classical methods.}
\label{fig:pyrazine_SPEED}
\end{figure}
An interesting comparison between Ehrenfest dynamics and SPEED can be made within the quadratic vibronic coupling Hamiltonian model, in which all diabatic PESs and all diabatic couplings between them are at most quadratic functions of the nuclear coordinates.
As follows from Eq.~(\ref{eq:Veff}), $\mathbf{V}_{\mathrm{eff}}(q,t)=\mathbf{V}(q)$ and Ehrenfest dynamics and SPEED are equivalent in this model, which is often used to represent real molecules.
Despite the harmonicity of the diabatic PESs, it is important to note that their adiabatic counterparts, obtained by diagonalization of the diabatic potential energy matrix at various nuclear configurations, is more flexible and can represent complex situations such as conical intersections between different electronic states. This simple observation explains the strength of an effective quadratic approximation of the potential in the diabatic representation. However, while adiabatic PESs can be readily computed using various electronic structure methods, constructing the potential in the diabatic representation typically requires considerably more effort. For this reason, applying SPEED to the quadratic vibronic coupling Hamiltonian model serves primarily as a proof of concept to showcase its accuracy. The efficiency is demonstrated in Sec.~\ref{sec:furan}, where SPEED is combined with an electronic structure method that provides the diabatic PESs \textit{ab initio}.

We begin by evaluating the performance of 1024 SPEED trajectories in describing the time-dependent population of the second excited state $S_{2}(\pi\pi^{*})$ of pyrazine, represented by a three-dimensional quadratic vibronic coupling Hamiltonian model~\cite{Woywod_Werner:1994}, following a UV excitation from the ground electronic and vibrational state. For comparison, we also propagated 1024 Ehrenfest trajectories, a single Ehrenfest trajectory, a thawed Gaussian Ehrenfest trajectory~\cite{Vanicek:2024,Scheidegger_Vanicek:2024}, and the exact quantum dynamics obtained with the second-order split-operator method~\cite{book_Tannor:2007}.
The initial positions and momenta for multi-trajectory methods are sampled from the Wigner distribution of the initial Gaussian nuclear wavepacket employed in the exact quantum dynamics.
We recently showed that a single Ehrenfest trajectory fails to describe diabatic transitions when the coupled electronic states belong to different irreducible representations of the molecular point group~\cite{Scheidegger_Vanicek:2024}. This limitation is also observed with the thawed Gaussian Ehrenfest dynamics, a method that combines single-trajectory Ehrenfest dynamics with semiclassical Gaussian wavepacket  description of nuclei~\cite{Vanicek:2024}.
For this reason, the electronic population of the second excited state, shown in Fig.~\ref{fig:pyrazine_SPEED}, remains constant for both single-trajectory methods. In contrast, Ehrenfest dynamics based on 1024 trajectories agrees with the exact quantum result almost perfectly for the first 800 a.u. and at least qualitatively for the remainder of the simulation. For nearly the cost of a single Ehrenfest trajectory, as a single reference Hessian obviously suffices in this case, SPEED offers the same accuracy as Ehrenfest dynamics.

\section{Adsorption on solid surface}
\label{sec:adsorption}
Atomic adsorption on a metallic surface illustrates well the limits of the Born--Oppenheimer approximation, as many electronic states must be included to accurately describe electronic friction induced by a dense manifold of electronic states~\cite{Dou_Subotnik:2018}. Accounting for these states requires a large computational overhead, and, for this reason, several friction theories have been developed to avoid directly addressing electron dynamics\cite{Persson_Hellsing:1982,Head-Gordon_Tully:1995,Dou_Subotnik:2016}. However, these approaches can fail dramatically when a single electron of the surface plays a major role in the dynamics, for example by transferring to the adsorbed molecule~\cite{Bartels_Wodtke:2011}. Such effects highlight the need for dynamic methods that can explicitly treat multiple closely spaced energy levels.
To address this, Tully and coworkers proposed the independent-electron surface hopping (IESH) algorithm~\cite{Shenvi_Tully:2009,Kruger_Schafer:2015}. This variant of the fewest-switches surface hopping method can handle an immense number of electronic states ($\approx10^{11}$) by assuming noninteracting electrons, thus replacing the conventional propagation of a single many-electron wavefunction with the propagation of an independent electronic wavefunction for each electron. When the approximations inherent to friction theories and IESH become inadequate, Ehrenfest dynamics offers an alternative, albeit at a significantly higher overhead---one that SPEED can reduce.

\begin{figure}
\centering
\includegraphics[width=\columnwidth]{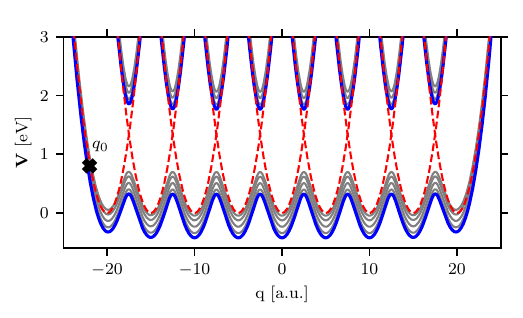}
\caption{Potential energy surfaces for the modeling of chemisorption of an atom on a metallic surface. The system is defined in the diabatic representation (dashed red) with horizontally shifted harmonic oscillators, each of which contains 5 states,  and constant diabatic couplings between them. The corresponding adiabatic curves are shown with solid gray lines, except for the ground state within a layer of closely lying electronic states, which is shown with a solid blue line. The initial position $q_{0}$ of the adsorbed atom is indicated by a black cross.}
\label{fig:adsorption_system}
\end{figure}
To evaluate the performance of SPEED in describing the chemisorption of an atom on a metallic surface, we employed the system proposed by Ryabinkin and Izmaylov~\cite{Ryabinkin_Izmaylov:2017a}. In their original paper, the authors demonstrated the improved accuracy of using collective electronic variables over friction theory to describe the time-dependent position of the adsorbed atom. As a reference benchmark, they used a simplified variant of single-trajectory Ehrenfest dynamics that neglects some contributions from the adiabatic excited states.
The model can be parameterized to represent various scenarios, from an insulator to a metallic surface. In one particularly challenging case for friction theory, this approximation incorrectly predicted that the adsorbed atom moving across the surface would abruptly change direction, whereas methods using a sufficient number of collective variables closely aligned with simplified Ehrenfest dynamics.
Using the parameters of this specific case, we observe that the expectation values of the position and potential energy obtained from a single Ehrenfest trajectory (without simplification) diverge significantly from the exact results (see Figs.~\ref{fig:adsorption_system_5layers_original} and \ref{fig:adsorption_pos_pot_5layers_original}). This observation highlights the necessity to go beyond single-trajectory Ehrenfest dynamics and other more approximate methods when computational resources are not a limiting factor.

Now, let us turn to a more realistic parameterization of the system that represents chemisorption on a metallic surface. The model, depicted in Fig.~\ref{fig:adsorption_system}, is one-dimensional and consists of nine primary diabatic harmonic oscillators, each horizontally displaced by 5 a.u., corresponding to the distance between the atoms forming the solid surface. Each harmonic potential is then duplicated four times and very slightly shifted horizontally ($\Delta q \approx 10^{-4}$ a.u.).
In the adiabatic representation, this quasi-degeneracy leads to closely spaced parallel electronic states arranged in layers, characteristic of a metallic surface, and ideal for Ehrenfest dynamics~\cite{Curchod_Tavernelli:2013,Loaiza_Izmaylov:2018}. To be able to perform exact quantum dynamics, we reduced the number of electronic states per layer from 10 to 5, compared to Ref.~\onlinecite{Ryabinkin_Izmaylov:2017a}.
The couplings between the diabatic PESs are constant. Consequently, all elements of the $45\times 45$ diabatic potential energy matrix are at most quadratic functions of the nuclear coordinate, rendering SPEED and Ehrenfest dynamics equivalent in this system.

\begin{figure}
\centering
\includegraphics[width=\columnwidth]{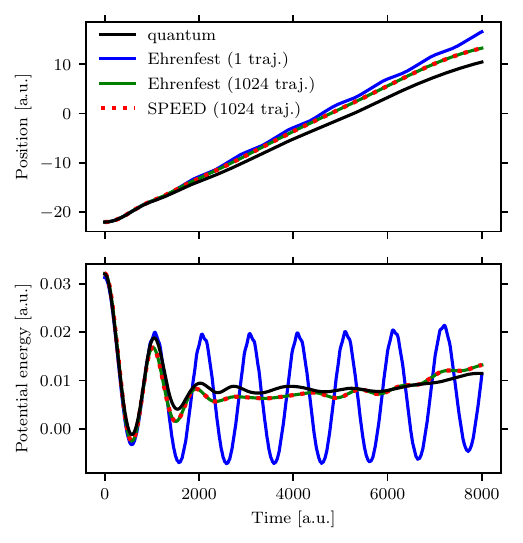}
\caption{Expectation value of position (top) and of the potential energy (bottom) of an atom chemisorbed on a metallic surface. Comparison of exact quantum dynamics with various mixed quantum-classical methods.}
\label{fig:adsorption_pos_pot}
\end{figure}

Two important aspects of chemisorption are the location of the adsorbed atom and the binding affinity between the surface and the atom. For this reason, we chose to analyze the expectation values of the position and potential energy. The top panel of Fig.~\ref{fig:adsorption_pos_pot} shows the position of the adsorbed atom evaluated exactly, with one Ehrenfest trajectory, with 1024 Ehrenfest trajectories, and with 1024 SPEED trajectories. The initial nuclear wavefunction for the exact propagation and sampling of the initial conditions for multi-trajectory methods is a Gaussian centered at $q_{0}=-22$ a.u., with zero momentum, and width corresponding to the vibrational ground state of any one of the diabatic harmonic oscillators.
All methods qualitatively agree on the expectation value of the position, but a single Ehrenfest trajectory deviates faster from the exact result than do the multi-trajectory methods.
We observe that more accurate propagation methods lead to a slower evolution of the adsorbed atom on the surface, likely due to a better incorporation of energy redistribution in more advanced methods. This suggests that chemisorption on a metallic surface requires not only a precise representation of the electronic states but also an accurate treatment of nuclear dynamics.

The expectation value of the potential energy, shown in the bottom panel of Fig.~\ref{fig:adsorption_pos_pot}, displays an initial perfect agreement between all methods. However, on a longer timescale, a single Ehrenfest trajectory proves insufficient, as the potential energy evaluated with this method continues to oscillate, whereas the exact result predicts stabilization. The nonphysical behavior of single-trajectory Ehrenfest dynamics can be understood from the bumpy shape of the ground adiabatic electronic surface in Fig.~\ref{fig:adsorption_system}, and the single Ehrenfest trajectory being represented by a classical particle.
In contrast, the full quantum nuclear wavefunction can spread out, which tends to damp the oscillations in the expectation value of the potential energy. This characteristic also applies to the swarm of trajectories propagated using Ehrenfest dynamics or SPEED, both of which demonstrate excellent agreement with the exact reference for the full simulation.

\section{Ab initio evaluation of hole transfer in furan dimer}
\label{sec:furan}
The diabatic PESs are often much simpler than their adiabatic counterparts. This is because diabatic PESs typically preserve certain physical properties of the system, which leads to their smoother behavior. In contrast, adiabatic PESs are more complex in shape, as they are determined by the eigenvalues of the time-independent electronic Schr\"odinger equation at various nuclear configurations and ordered by energy. However, identifying a meaningful conserved property to construct a PES in the diabatic representation is often a challenge. A more common approach consists of constructing effective diabatic PESs and couplings using power series expansions around a reference geometry. The coefficients in the expansion are then determined by the condition that diagonalizing the diabatic matrix recovers the corresponding adiabatic energies~\cite{Evenhuis_Martinez:2011}.

\begin{figure}
\centering
\includegraphics[width=\columnwidth]{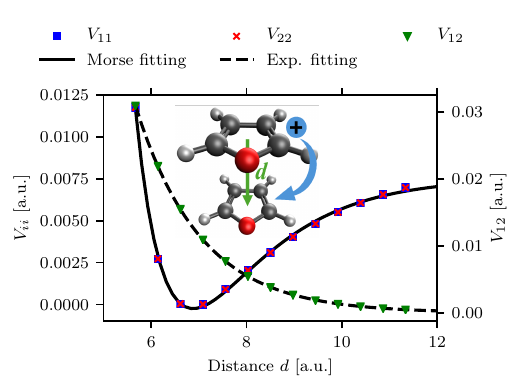}
\caption{Coupled Morse system representing hole transfer between two furan monomers. The degenerate diabatic PESs (blue squares and red crosses) and their diabatic coupling (green triangles) have been evaluated \textit{ab initio} at various monomers distances, before being fitted by Morse (solid) and exponential (dashed) curves, respectively.}
\label{fig:furane_dimer_system}
\end{figure}

In certain applications, the conserved property along a diabatic PES is evident. This is particularly true for charge (hole or electron) transfer between two distinct moieties, identified as either the donor or acceptor. One diabatic curve is computed by constraining the charge to remain on the donor, while the other curve is obtained by localizing the charge on the acceptor fragment. Markland and coworkers recently proposed an electronic structure method that constructs absolutely localized molecular orbitals (ALMO) and uses multistate density functional theory (MSDFT) to evaluate the diabatic potential energy and diabatic coupling for charge transfer processes. The resulting scheme, called ALMO(MSDFT2)~\cite{Mao_Markland:2019}, was implemented in the Q-Chem package~\cite{Shao_Head-Gordon:2015}. As mentioned by the authors, the efficiency of the method opens the door for the description of charge transfer using \textit{ab initio} quantum dynamics in the diabatic representation, which conveniently avoids the complexities of the adiabatic representation associated with the geometric phase~\cite{Ryabinkin_Izmaylov:2014} and the divergence of derivative couplings at conical intersections. Motivated by this conclusion, we constructed global diabatic potential energy curves of the furan dimer, one of the compounds used to benchmark ALMO(MSDFT2), at the $\omega$B97XD~\cite{Chai_Head-Gordon:2008,Chai_Head-Gordon:2008a}/6-31+G(d) level of theory. The system, represented in Fig.~\ref{fig:furane_dimer_system}, describes the transfer of a hole from one monomer to the other. The dimer consists of two identical moieties, resulting in globally degenerate diabatic curves when plotted as a function of the monomers distance. Consequently, the well-known Landau-Zener transition probability~\cite{Zener:1932,Zhu_Nakamura:1995} is not defined for this system because the analytical formula involves a division by the difference of diabatic forces, which is zero here. Therefore, evaluating the charge transfer probability requires an explicit description of the dynamics.
Although straightforward to implement, an on-the-fly evaluation of the potential energy information needed in trajectory-based methods would require numerical differentiation to obtain the gradients and Hessians of the PESs and their diabatic couplings at each time step, as the analytical derivatives of the energy have not yet been implemented in Q-Chem for ALMO(MSDFT2). For this reason, we numerically evaluated the diabatic potential energy matrix at various separations between monomers and then used Morse and exponential fittings to define analytical diabatic potential energy curves that can be applied to both trajectory-based methods and exact quantum dynamics. The fitting parameters can be found in Sec.~\ref{sec:SI_morse_exp_parameters} of the Supplemental Material.

We simulated the hole transfer dynamics between furan monomers initially placed at a sufficient distance for the coupling to be zero. Initially, only one of the diabatic states is occupied, and the nuclear wavepacket is a Gaussian centered at $q_{0}=20$ a.u. and with a width corresponding to the vibrational ground state of the stable neutral dimer. To estimate the effect of temperature on the charge transfer rate, the initial momentum of the center of the nuclear wavepacket is set to
%
\begin{equation}
    p(T)=\mu \langle v_{\mathrm{rel}} \rangle = \mu\sqrt{\frac{8k_{B}T}{\pi\mu}},
\end{equation}
where $\langle v_{\mathrm{rel}} \rangle$ is the mean relative velocity of a Maxwell--Boltzmann distribution at temperature $T$, $k_{B}$ is the Boltzmann constant, $\mu = M/2$ is the reduced mass of the dimer, and $M$ is the mass of one monomer.

\begin{figure}
\centering
\includegraphics[width=\columnwidth]{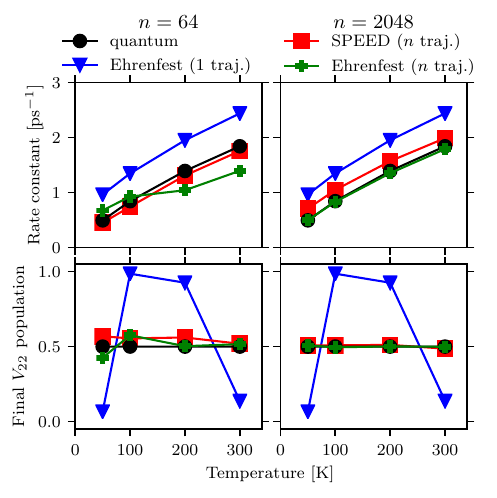}
\caption{Temperature dependence of the hole transfer rate constant (top) and final population of the second diabatic state (bottom) evaluated exactly (black discs), with a single (blue triangles) or multiple (green crosses) Ehrenfest dynamics and with SPEED (red squares).}
\label{fig:rate_constant_pop}
\end{figure}
The electronic population itself is not an observable. However, in this case, the time-dependent occupation of the diabatic states directly indicates the distribution of the charge between the two monomers. In particular, the most important features are the early time of the charge transfer, for which a rate constant can be defined (details of its evaluation are provided in Sec.~\ref{sec:SI_rate_constant} of the Supplemental Material), and the final charge distribution between the two furan moieties after the collision event.
Figure~\ref{fig:rate_constant_pop} presents the results obtained with SPEED and Ehrenfest dynamics at different temperatures using 64 and 2048 trajectories, along with the exact quantum reference and the result obtained using a single Ehrenfest trajectory. With only 64 trajectories, SPEED accurately captures the expected increase in the rate constant at higher temperatures, whereas Ehrenfest dynamics behaves less predictably.
The trend is also captured with a single Ehrenfest trajectory, but a large overestimation of the rate is observed. With 1024 trajectories, both Ehrenfest dynamics and SPEED are fully converged and capture the trend, but SPEED somewhat overestimates the rate of hole transfer, whereas Ehrenfest dynamics converges to a value in excellent agreement with the exact results.

The lower panels show that exact quantum dynamics yields a final population of one-half for both diabatic states across all temperatures, indicating an equal probability of the charge being localized on either monomer after the collision. Whereas a single Ehrenfest trajectory completely fails to describe this effect, 64 Ehrenfest and 64 SPEED trajectories provide a reasonable estimate of the charge distribution, although a minor statistical deviation from the exact result persists. When fully converged, both multi-trajectory methods accurately recover the 1:1 ratio.
We want to remind the reader, however, that Ehrenfest dynamics does not satisfy detailed balance~\cite{Parandekar_Tully:2006}, which generally limits its ability to accurately describe equilibrium properties. In particular, electronic populations at equilibrium diverge dramatically from the Boltzmann distribution as the energy gap between the electronic states increases~\cite{Parandekar_Tully:2005}.

\section{Limitations of SPEED}
\label{sec:limitation_SPEED}
Finally, let us discuss the limitations of our approach. In order for SPEED to produce accurate results, it is necessary that \emph{both} Ehrenfest dynamics itself is accurate and the diabatic PESs and couplings are only weakly anharmonic.

The first condition implies that SPEED inherits all the drawbacks of standard Ehrenfest dynamics, which is a locally mean-field mixed quantum-classical method. The limitations of Ehrenfest dynamics are well known and include the lack of interference between different trajectories due to the classical treatment of the nuclear motion.
In addition, after leaving a region of strong nonadiabatic coupling, each Ehrenfest trajectory typically continues to evolve on an average electronic surface, rather than collapsing onto a single surface~\cite{Zimmermann_Vanicek:2012,Curchod_Tavernelli:2013}.
In the nonadiabatic field method developed by Liu and coworkers~\cite{Wu_Liu:2024}, this nonphysical effect is corrected by defining the nuclear force as a sum of two terms. The first one is the usual nonadiabatic nuclear force, but the second one corresponds to the adiabatic force from a single adiabatic electronic state, rather than from the weighted average over all electronic states.

\begin{figure}
\centering
\includegraphics[width=\columnwidth]{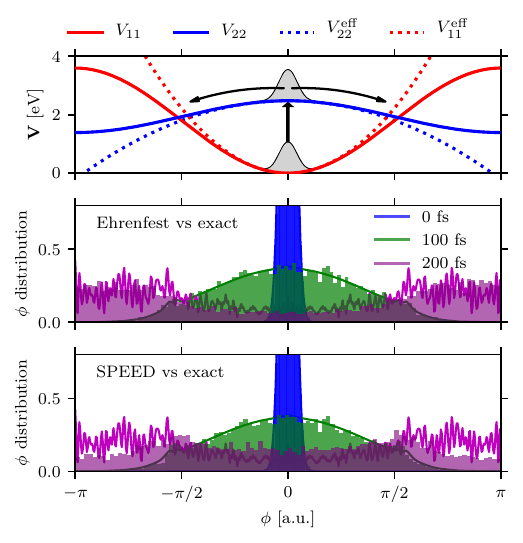}
\caption{Top: Potential energy profile of retinal along the isomerization coordinate $\phi$, with solid lines representing the diabatic curves and dotted lines indicating the diagonal elements of the effective potential energy matrix used by SPEED. Excitation of the initial nuclear wavepacket and the following dynamics are represented schematically. Middle: Nuclear density obtained from the exact nuclear wavepacket (solid lines) or from the histogram of final coordinates of the 8192 Ehrenfest trajectories at various times. Bottom: Same as the middle panel, except that the histogram is for 8192 SPEED trajectories.}
\label{fig:retinal_system}
\end{figure}

Because the limitations of Ehrenfest dynamics have been studied extensively, in this final section we focus on the second condition, i.e., on the limitations of SPEED beyond those of Ehrenfest dynamics. In particular, we will show an example where Ehrenfest dynamics remains accurate, while strong anharmonicity results in significant errors when using SPEED.
We use the two-dimensional diabatic model of retinal proposed by Hahn and Stock~\cite{Hahn_Stock:2000}, in which one coordinate is harmonic and collectively represents all vibrations coupling the electronic ground and first excited states, while the second coordinate $\phi$ represents the torsional angle associated to the cis-trans isomerization. A cut of the potential along the isomerization coordinate is shown in the top panel of Fig.~\ref{fig:retinal_system}, together with a schematic representation of the photoexcitation from the stable cis-isomer and ensuing dynamics. Due to the local maximum at the Franck--Condon point on the excited-state surface, the quantum wavepacket splits and moves in the directions of the two adjacent local minima, both associated with the trans-isomer. To account for the 2$\pi$-periodicity of the isomerization coordinate, periodic boundary conditions are imposed at $\phi=-\pi$ and $\phi=\pi$ to maintain continuity. This ensures that trajectories crossing one boundary emerge on the opposite side. The initial effective potential felt by the SPEED trajectories is represented by dotted lines.

To assess the ability of Ehrenfest dynamics and SPEED in describing the isomerization process, we evaluate the time-dependent probability distribution of the isomerization coordinate $\phi$ (middle and bottom panels of Fig.~\ref{fig:retinal_system}).
For multi-trajectory methods, the nuclear density along $\phi$ is represented using a histogram of the coordinates of all trajectories, while the exact one is obtained by integrating the full molecular density over the other nuclear coordinate. During the first 100 fs, both Ehrenfest dynamics and SPEED remain in excellent agreement with exact quantum dynamics, and both methods capture the expected spread of the initial nuclear Gaussian wavepacket. Over a longer time scale, Ehrenfest dynamics successfully describes the formation of the trans-isomer ($\phi=\pm\pi$) within 200 fs, as observed experimentally~\cite{Peteanu_Shank:1993} and later reproduced in quantum simulations~\cite{Hahn_Stock:2000}.

In contrast, SPEED fails to accurately capture the isomerization process (bottom panel of Fig.~\ref{fig:retinal_system}). This is because, when SPEED is fully converged (i.e., in the limit of an infinite number of trajectories), by symmetry the effective potential (\ref{eq:Veff}) does not evolve over time and remains centered at $\phi_{\mathrm{avg},t}=0$, around which the nuclear trajectories are distributed equally.
Consequently, the effective PESs used in SPEED are very inaccurate near \(\phi = \pm\pi\). However, because of the periodic boundary conditions, local minima still appear at these points. Without such boundary conditions---for instance, when the excited-state surface exhibits a double-well structure~\cite{Lehr_Graham:2020,Vester_Kuleff:2023}---the issue becomes even more severe, as the SPEED trajectories would simply spread indefinitely.

\begin{figure}
\centering
\includegraphics[width=\columnwidth]{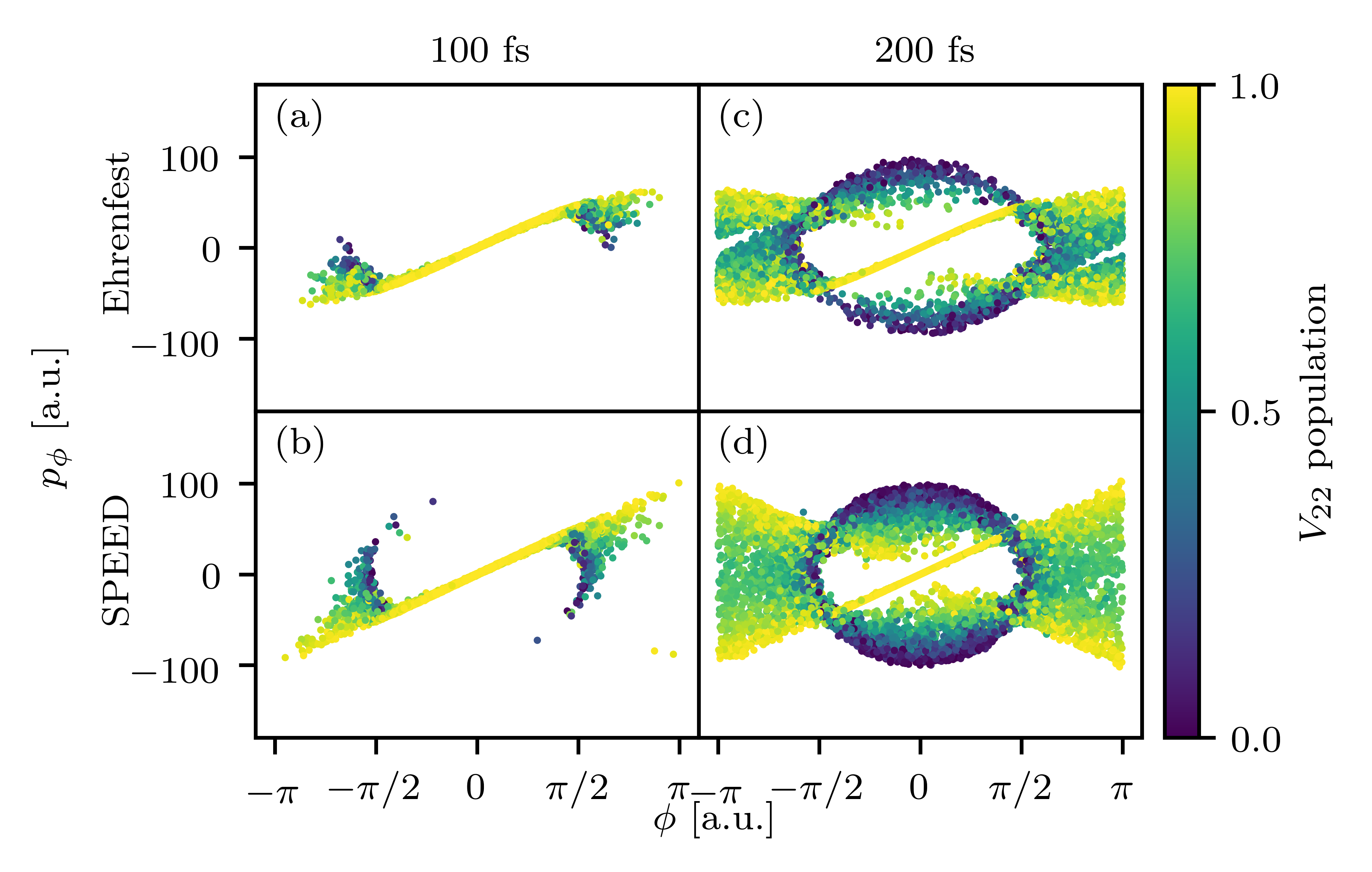}
\caption{Nuclear coordinates of 8192 SPEED and 8192 Ehrenfest trajectories in the phase space associated with the isomerization coordinate of retinal. The distributions of positions and momenta are shown at $t=100$ fs and $t=200$ fs, with each phase-space point colored according to the electronic population of the diabatic excited state.}
\label{fig:retinal_ED_SPEED}
\end{figure}
The transfer of electronic population and important differences between SPEED and Ehrenfest dynamics are more apparent in Fig.~\ref{fig:retinal_ED_SPEED}, which presents a scatter plot of the nuclear coordinates in the phase space associated with the isomerization coordinate. The marker colors indicate the fraction of electronic population associated with the diabatic excited state.
The nuclear trajectories forming a yellow diagonal in panels (a) and (b) illustrate the initial spread of the Gaussian wavepacket (mostly) on the excited-state surface within the first 100 fs along both the position and momentum coordinates associated with $\phi$. At each end of this diagonal, the trajectories split into two distinct paths. One path leads to the formation of the trans-isomer while remaining in the diabatic excited state, whereas the other enables relaxation to the electronic ground state, resulting in harmonic motion of the associated trajectories around the initial configuration. This second effect is particularly visible with SPEED, where electronic transitions are more pronounced.
To justify this interpretation, see Fig.~\ref{fig:retinal_ED_SPEED_no_coupl}, which is an analog of Fig.~\ref{fig:retinal_ED_SPEED}, but where the couplings between the surfaces are ignored.

The overestimation of the probability to regenerate the cis-isomer in its electronic ground state with SPEED arises from the steeper topology of the crossing between $V_{22}^{\mathrm{eff}}$ and $V_{11}^{\mathrm{eff}}$ relative to that between $V_{22}$ and $V_{11}$. This explanation is consistent with studies showing that sloped conical intersections enhance photostability by facilitating rapid return to the initial configuration in its electronic ground state~\cite{Hall_Robb:2006,Tuna_Domcke:2016}.
Another notable difference is that SPEED tends to overestimate the momentum of trajectories evolving on the diabatic excited-state surface. 
While after 200 fs most Ehrenfest trajectories accumulate near $\phi=\pm\pi$, indicating completion of the isomerization process, a significant fraction of SPEED-propagated trajectories on the diabatic excited state gained sufficient momentum to escape the local minimum associated with the photoproduct and reform the initial configuration while remaining on the diabatic excited state.

\section{Conclusion and outlook}
\label{sec:conclusion}
To conclude, we have introduced SPEED, a variation of Ehrenfest dynamics that employs a single, time-dependent quadratic effective potential to propagate all nuclear trajectories.
As a result, the computational cost remains mostly independent of the number of trajectories when evaluating the potential energy information is the limiting factor, as in on-the-fly \textit{ab initio} applications. We demonstrated the equivalence of standard Ehrenfest dynamics and our new approach in two realistic models describing the internal conversion in pyrazine and chemisorption of an atom on a solid surface. This equivalence arises because the diabatic PESs and their couplings are, at most, quadratic functions of the nuclear coordinates.

A diabatic potential is typically not obtained directly but is instead constructed from adiabatic \textit{ab initio} PESs. Yet, some electronic structure methods can compute the diabatic potential energy matrix directly by applying physically meaningful constraints. The recently developed ALMO(MSDFT2) method~\cite{Mao_Markland:2019}, which describes charge transfer between diabatic electronic states, belongs to this category, opening the door to on-the-fly quantum dynamics in the diabatic representation. Using ALMO(MSDFT2), we evaluated the rate constant for hole transfer between two furan molecules and analyzed the final charge distribution after the collision at various temperatures. We find that SPEED requires fewer trajectories than standard Ehrenfest dynamics to capture the increase in the rate constant at higher temperatures.
However, a systematic overestimation of the rate is observed. At convergence, both Ehrenfest dynamics and SPEED accurately predict the charge distribution between the two monomers.
The faster convergence rate of SPEED can be attributed to its simpler dynamics, as all trajectories are confined within the same multistate harmonic potential. However, as highlighted in the retinal example, SPEED may converge to an incorrect result when the diabatic PESs are particularly anharmonic.

We conclude that our new approach provides a reliable qualitative description of nonadiabatic dynamics across a wide range of molecular systems. Assuming diabatic PESs can be obtained---for instance, through on-the-fly diabatization procedures~\cite{Choi_Martinez:2004,Varga_Truhlar:2018,Richings_Habershon:2020}---SPEED offers significant efficiency gains over Ehrenfest dynamics, which can be particularly advantageous for large systems.
SPEED can be improved by employing alternative effective potentials to improve efficiency or accuracy. This can be achieved, for instance, by using a single reference Hessian throughout the propagation or by increasing the order of the effective potential to better capture nonlocal effects.
Additionally, the ``Single Potential Evaluation'' strategy can be combined with more advanced multi-trajectory methods than Ehrenfest dynamics, for example with the multi-configurational Ehrenfest method developed by Shalashilin. This approach, which is in principle exact, employs nuclear frozen Gaussian wavepackets following Ehrenfest trajectories as a basis to variationally determine the optimal solution~\cite{Shalashilin:2009,Saita_Shalashilin:2012} and capture correlations between trajectories that are neglected in standard Ehrenfest dynamics.

\section*{Acknowledgments}
The authors acknowledge financial support from the Swiss National Science Foundation through the National Center of Competence in Research MUST (Molecular Ultrafast Science and Technology) and from the European Research Council (ERC) under the European Union's Horizon 2020 research and innovation program (Grant Agreement No. 683069--MOLEQULE).

\bibliographystyle{apsrev4-2}
\bibliography{biblio61,additions_SPEED}

\clearpage
\widetext
\begin{center}
\textbf{\large Supplemental Material for ``Ehrenfest dynamics accelerated with SPEED''}
\end{center}
\setcounter{equation}{0}
\setcounter{figure}{0}
\setcounter{table}{0}
\setcounter{section}{0}
\setcounter{page}{1}
\makeatletter
\renewcommand{\thefigure}{S\arabic{figure}}
\renewcommand{\bibnumfmt}[1]{[S#1]}

\section{Computational overhead comparison between Ehrenfest dynamics and SPEED}
\label{sec:SI_computational_cost}
In Ehrenfest dynamics, the anharmonic (e.g., \textit{ab initio}) energy and force must be calculated for each trajectory. In contrast, SPEED requires only a single evaluation of the energy, force, but also the Hessian. This raises the question: Under what conditions does SPEED offer an efficiency advantage? To give a rough estimate, we compare the cost given by the number of unique scalars that must be computed to describe the potential in a simulation treating $S$ electronic states, $D$ nuclear dimensions, and $N_{\mathrm{traj}}$ trajectories.

This cost is
\begin{equation}
\label{eq:cost_ED}
    C_{\mathrm{Ehrenfest}}=\frac{1}{2}S(S+1)(1+D)N_{\mathrm{traj}}
\end{equation}
for Ehrenfest dynamics and
\begin{equation}
\label{eq:cost_SPEED}
    C_{\mathrm{SPEED}}=\frac{1}{2}S(S+1)[1+D+D(D+1)/2]
\end{equation}
for SPEED. Equating Eqs.~(\ref{eq:cost_ED}) and (\ref{eq:cost_SPEED}) yields $N_{\mathrm{traj}}:=1+D/2$. Therefore, SPEED remains more efficient than Ehrenfest dynamics for systems where the number of trajectories required for convergence is larger than one half of the number of nuclear degrees of freedom.

\section{Chemisorption of an atom on a metallic surface}
\label{sec:SI_chemisorption}
\begin{figure}[H]
\centering
\includegraphics[width=0.75\columnwidth]{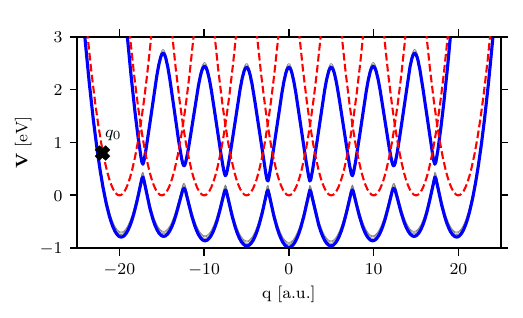}
\caption{Potential energy surfaces for the chemisorption of an atom on a metallic surface. The system is defined in the diabatic representation with horizontally shifted harmonic oscillators (dashed red) and constant couplings between them. The corresponding adiabatic curves are shown with solid blue (ground state within a layer) and gray lines. The initial position $q_{0}$ of the adsorbed atom is indicated by a black cross.}
\label{fig:adsorption_system_5layers_original}
\end{figure}

\begin{figure}[H]
\centering
\includegraphics[width=0.75\columnwidth]{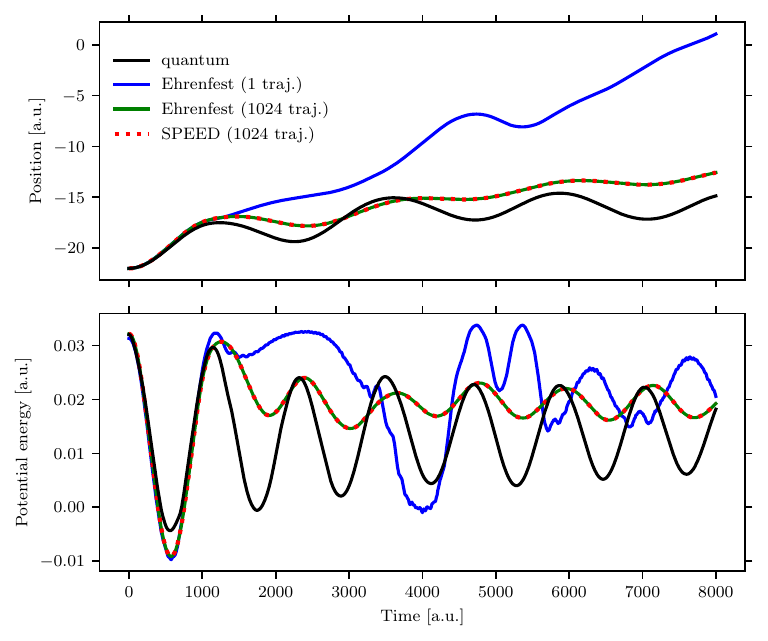}
\caption{Expectation values of position (top) and of the potential energy (bottom) of an atom chemisorbed on a one-dimensional chain of atoms. Comparison of exact quantum dynamics with various mixed quantum-classical methods. The system parameterization for these simulations was selected in Ref.~\onlinecite{Ryabinkin_Izmailov:2017} to emphasize the limitations of friction theory. This explains why the result from a single Ehrenfest trajectory deviates here significantly more from the exact result than in Fig.~\ref{fig:adsorption_pos_pot}.}
\label{fig:adsorption_pos_pot_5layers_original}
\end{figure}

The model proposed in Ref.~\onlinecite{Ryabinkin_Izmaylov:2017a} was also parameterized to simulate a case (see Fig.~\ref{fig:adsorption_system_5layers_original}) where friction theory would qualitatively fail by incorrectly predicting a sudden change of direction of the adsorbed atom. The performance of SPEED in this system, compared to single- and multi-trajectory Ehrenfest dynamics, as well as to exact quantum dynamics, is shown in Fig.~\ref{fig:adsorption_pos_pot_5layers_original}. A single Ehrenfest trajectory poorly describes the position and potential energy of the adsorbed atom. In contrast, both multi-trajectory methods provide a good qualitative picture for both observables.

\section{Charge transfer in furan dimer}
\subsection{Fitted Morse and exponential parameters}
\label{sec:SI_morse_exp_parameters}
The degenerate diabatic potential energy curves of the furan dimer model, depicted in Fig.~\ref{fig:furane_dimer_system}, are fitted with a Morse function
\begin{equation}
\label{eq:Morse_fit}
    V^{\mathrm{Morse}}(d) = V_{0}^{\mathrm{Morse}} + c (1 - e^{{-a(d-d_0)}})^{2}.
\end{equation}
The parameters, in atomic units, are $V_{0}^{\mathrm{Morse}}=-0.000252592$, $c=0.00780249$, $a=0.665027$ and $d_{0}=6.88046$. The diabatic couplings approximately follow an exponential function
\begin{equation}
\label{eq:exp_fit}
    V^{\mathrm{exp}}(d) = V^{\mathrm{exp}}_{0}e^{-\lambda d}
\end{equation}
with parameters $V^{\mathrm{exp}}_{0}= 2.01656$ and $\lambda=0.736526$.

\subsection{Evaluation of the rate constant}
\label{sec:SI_rate_constant}
\begin{figure}[H]
\centering
\includegraphics[width=0.75\columnwidth]{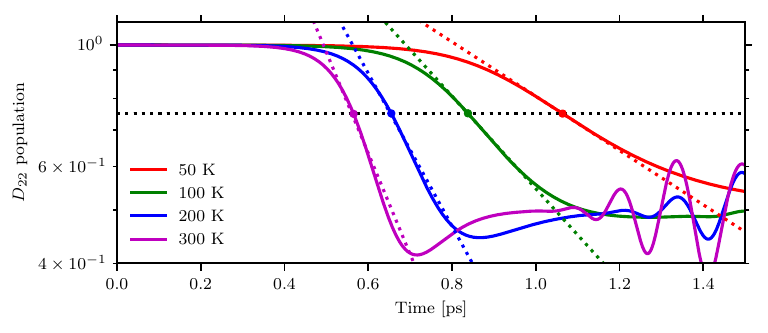}
\caption{Time dependence of the population of the second diabatic state of the furan dimer evaluated with exact quantum dynamics at various temperatures (solid lines). On a logarithmic scale, the population decay is approximately linear around the point where the population is 0.75 and the rate constant is defined to be the slope of a linear fit.}
\label{fig:rate_constant_evaluation}
\end{figure}

The two furan monomers are initially at a sufficiently large distance for the diabatic coupling to be effectively zero. As the distance between the donor and acceptor decreases, charge transfer occurs from one moiety to the other. As an example, Fig.~\ref{fig:rate_constant_evaluation} shows the exact quantum population of the first excited state as a function of time at different temperatures. Although some oscillations can be observed after the initial decay before stabilizing at the value of 1/2, the population dynamics essentially follows a sigmoid curve for which a rate constant can be defined by an exponential fitting at the point where the electronic population is 0.75.

\section{Retinal photoisomerization}
\label{sec:SI_retinal}
\subsection{Dynamics without diabatic coupling}
\begin{figure}[H]
\centering
\includegraphics[]{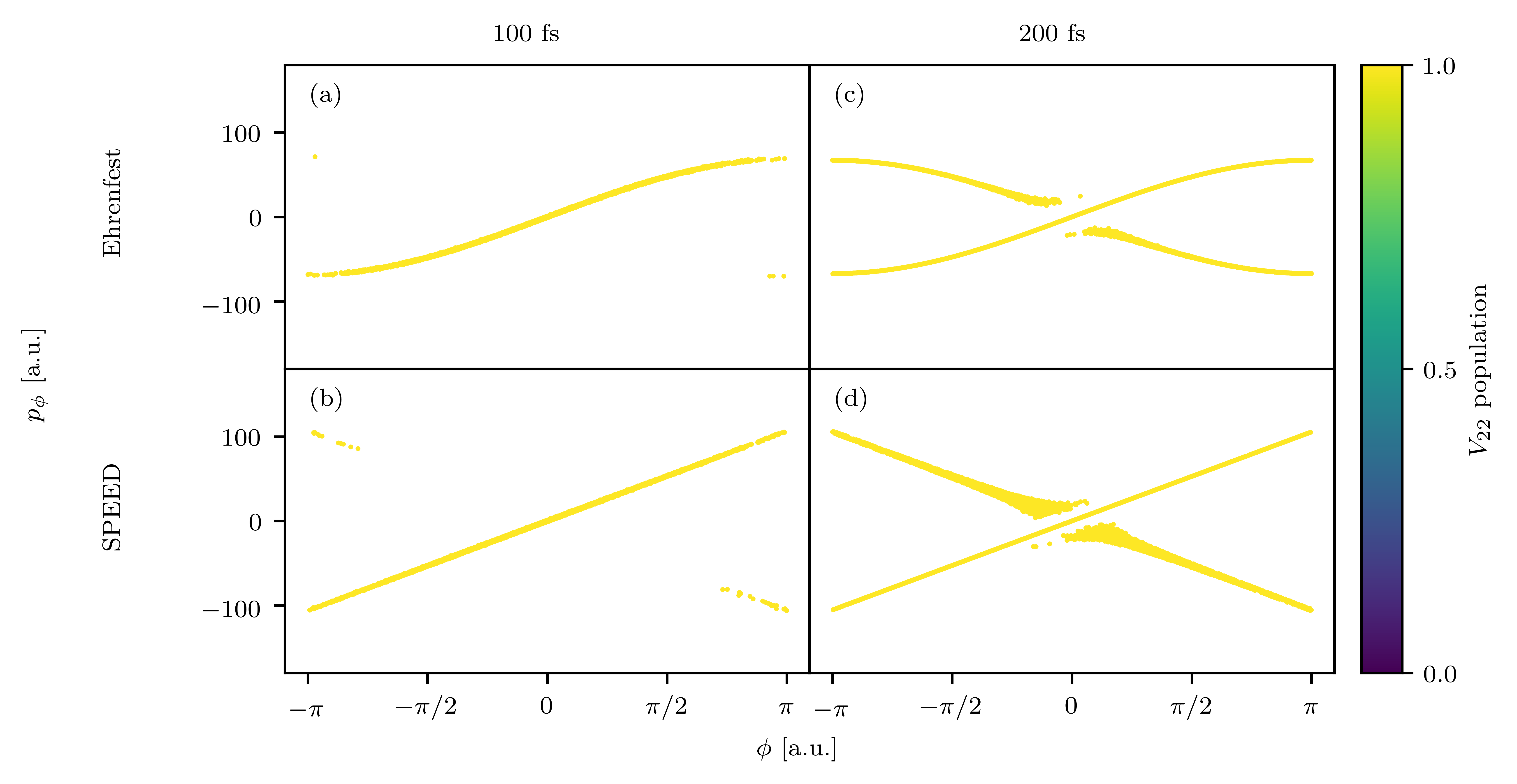}
\caption{Nuclear coordinates of 8192 SPEED and 8192 Ehrenfest trajectories in the phase space associated with the isomerization coordinate. The distributions of positions and momenta are shown at $t=100$ fs and $t=200$ fs, with each phase-space point colored according to the electronic population of the diabatic excited state. This figure is an analogue to Fig.~\ref{fig:retinal_ED_SPEED} of the main text, except that the diabatic couplings are neglected here. Therefore, population of the excited state remains constant and all points are yellow.}
\label{fig:retinal_ED_SPEED_no_coupl}
\end{figure}

In the diabatic representation, retinal isomerization occurs through the motion of the nuclear wavepacket on the excited-state diabatic potential surface. Figure~\ref{fig:retinal_ED_SPEED_no_coupl} compares Ehrenfest dynamics and SPEED in the retinal model, with the diabatic coupling artificially removed to better highlight the differences between the two methods on the excited-state surface. The inadequate representation of the potential in the trans-isomer region ($\phi\pm\pi$) with SPEED leads to an overestimation of the momentum after the photoproduct forms, causing the trajectories to prematurely return to the cis-configuration.

\end{document}